\documentclass{article}
\usepackage{graphicx}
\usepackage{amsmath}
\usepackage{amsfonts}
\usepackage{amssymb}
\newtheorem{theorem}{Theorem}

\newtheorem{definition}[theorem]{Definition}

\begin{document}

\title{Lightfront holography and area density of entropy associated with localization
on wedge-horizons}
\author{Bert Schroer\\present address: CBPF, Rua Dr. Xavier Sigaud 150, \\22290-180 Rio de Janeiro, Brazil\\email schroer@cbpf.br\\permanent address: Institut f\"{u}r Theoretische Physik\\FU-Berlin, Arnimallee 14, 14195 Berlin, Germany}
\date{June 2002 }
\maketitle
\begin{abstract}
It is shown that a suitably formulated algebraic lightfront holography (LFH),
in which the lightfront is viewed as the linear extension of the upper causal
horizon of a wedge region, is capable of overcoming the shortcomings of the
old lightfront quantization. The absence of transverse vacuum fluctuations
which this formalism reveals is responsible for an area (edge of the wedge)
-rearrangement of degrees of freedom which in turn leads to the notion of area
density of entropy for a ``split localization''. This area proportionality of
horizon associated entropy has to be compared to the volume dependence of
ordinary heat bath entropy. The desired limit, in which the split distance
vanishes and the localization on the horizon becomes sharp, can at most yield
a relative area density which measures the ratio of area densities for
different quantum matter. In order to obtain a normalized area density one
needs the unknown analog of a second fundamental law of thermodynamics 

for thermalization caused by vacuum fluctuation through localization on causal
horizons. This is similar to the role of the classical Gibbs form of that law
which relates 

Bekenstein's classical area formula with the Hawking quantum mechanism for
thermalization from black holes.

PACS: 11.10.-z, 11.30.-j, 11.55.-m
\end{abstract}

\section{Constructive Aims of Lightfront Holography}

Lightfront quantum field theory and the closely related $p\rightarrow\infty$
frame method have a long history. The large number of articles on this subject
(which started to appear at the beginning of the 70ies) may be separated into
two groups. On the one hand there are those papers whose aim is to show that
such concepts constitute a potentially useful enrichment of standard local
quantum physics \cite{Leut}\cite{DriesslerI}\cite{DriesslerII}, but there are
also innumerable attempts to use lightfront concepts as a starting point of
more free-floating ``effective'' approximation ideas in high-energy
phenomenology (notably for Bjorken scaling) whose relations to causal and
local quantum physics remained unclear or were simply not addressed.

As will become clear in the next section, where we will recall some of the old
ideas from a modern perspective, the old lightfront approach was severely
limited since in d=1+3 spacetime dimensions its canonical short distance
prerequisites are met only in the absence of interactions. Nevertheless
algebraic lightfront holography\footnote{The term ``holography'' was
introduced by 't Hooft \cite{H} for the description of his intuitive idea
about the organization of degrees of freedom in the presence of event horizons
for QFT in CST. The present setting of LFH is algebraic QFT (AQFT) in
Minkowski spacetime.} (LFH), which is the subject of this paper, may be viewed
as a revitalization of the old approach with new concepts. The aim of the old
approach (though never satisfactorily achieved) was the simplification of
dynamics by encoding some of its aspects into more sophisticated kinematics;
this is precisely what LFH aims to achieve, but this time without suffering
from short distance limitations which eliminate interactions and with the full
awareness of the locality issue and the problem of reconstruction of the
original theory (or family of original theories) which has (have) the same
holographic projection.

Whereas the old approach amounted to the lightfront restriction of pointlike
fields (which also caused the mentioned limitation), the LFH reprocesses the
original fields first into a net of algebras which by algebraic holography is
then converted into a net of operator algebras indexed by spacetime regions in
the lightfront. This net turns out to be a ``generalized'' chiral net (a
chiral net extended by a very specific vacuum-polarization-free transverse
quantum mechanics) with a 7-parametric vacuum symmetry group which corresponds
to a subgroup of the 10-parametric Poincar\'{e} group of the ambient theory.
It also possesses additional higher symmetries which originate from the
conformal covariance of the chiral LFH theory. The latter amount to
diffuse-acting automorphisms in the ambient theory (``fuzzy symmetries'').
This does not only include the rigid rotation which belongs to the Moebius
group with $L_{0}$ as its infinitesimal generator, but also all higher
diffeomorphism of the circle $Diff(S^{1})$.

It is important to realize that the absence of a direct relation between the
ambient fields $A$ and those $A_{LF}$ generating the lightfront net i.e.
\begin{equation}
A_{LF}(x)\neq A(x)|_{LF}%
\end{equation}
is the prize to pay for the simplification of \textit{interacting} quantum
field theories\footnote{In fact in the presence of strictly renormalizable
interactions, the right hand side is void of mathematical meaning (see
\ref{inter} in section 3).} in the algebraic LFH (where the left hand side is
determined by intermediate algebraic steps and not by restricting fields). The
inverse holography i.e. the classification of ambient theories which belong to
one LFH class (including the action of its 7-parametric symmetry group)
remains as the main unsolved problem.

Although the LFH does not accomplish dynamical miracles, it shifts the
separating line between kinematical substrate and dynamical actions in a
helpful way by placing more structure (as e.g. compared to the canonical
formalism) onto the kinematical side which is described by the holographic
projection. LFH also avoids having an ``artistic'' starting point as e.g.
canonical commutation relations or functional integral representations (which
the physical renormalized theory cannot maintain in the presence of
interactions, apart from d=1+1 polynomial scalar interactions). Instead it
fits well into the spirit of Wightman QFT or algebraic QFT (AQFT), where the
result obtained at the end of a computation should be (and are) in complete
mathematical harmony with the requirements at the start. Among the reasonably
easy structural consequences of LFH is the surface proportionality of
\textit{localization entropy} associated with a causal horizon. This will be
the subject of the previous to last section.

There are several papers which discuss the gain of scale symmetry through
restrictions of QFT to horizons using modular theory \cite{Sewell}%
\cite{Verch}\cite{GLRV} without addressing the problem of the transverse
direction which is the central point of the present work.

Although the connection between the local net on the lightfront and that on
the full ambient spacetime turns out to be quite nonlocal (in
contradistinction to the algebraic AdS-CQFT holography, which still preserves
many relative local aspects \cite{Rehren}), the modular localization approach
succeeds to formulate this holographic procedure (including an understanding
of its nonlocal relations to the original theory) in the rigorous setting of
local quantum physics. With these remarks on what LFH means in this paper (as
compared to many other meanings of ``holography'' in the recent literature),
we conclude our historical and general remarks and pass to a mathematical
description \footnote{The present work combines and supersedes previous
unpublished work of the author (hep-th 0106284, 0108203, 0111188) and
emphasizes different aspects of the subject published in \cite{parad}.}.

\section{Elementary facts on pointlike fields restricted to the lightfront}

For some elementary observations we now turn to the simple model of a d=1+1
massive free field
\begin{align}
A(x)  &  =\frac{1}{\sqrt{2\pi}}\int\left(  e^{-ipx}a(\theta)+e^{ipx}a^{\ast
}(\theta)\right)  d\theta\\
p  &  =m(ch\theta,sh\theta)\nonumber
\end{align}
where for convenience we use the momentum space rapidity description. In order
to get onto the light ray $x_{-}=t-x=0$ in such a way that $x_{+}=t+x$ remains
finite, we approach the $x_{+}>0$ horizon of the right wedge $W:$ $x>\left|
t\right|  $ by taking the $r\rightarrow0,\,\chi=\hat{\chi}-ln\frac{r}{r_{0}%
}\rightarrow\hat{\chi}$ $+\infty$ limit in the x-space rapidity
parametrization
\begin{align}
&  x=r(sh\chi,ch\chi),\,\,x\rightarrow(x_{-}=0,x_{+}\geq0,\text{
}finite)\label{lim}\\
&  A(x_{+},x_{-}\rightarrow0)\equiv A_{LF}(x_{+})=\frac{1}{\sqrt{2\pi}}%
\int\left(  e^{-ip_{-}x_{+}}a(\theta)+e^{ip_{-}x_{+}}a^{\ast}(\theta)\right)
d\theta\nonumber\\
&  =\frac{1}{\sqrt{2\pi}}\int\left(  e^{-ip_{-}x_{+}}a(p)+e^{ip_{-}x_{+}%
}a^{\ast}(p)\right)  \frac{dp}{\left|  p\right|  }\nonumber
\end{align}
where the last formula exposes the limiting $A_{LF}(x_{+})$ field as a chiral
conformal (gapless $P_{-}$ spectrum) field; the mass in the exponent
$p_{-}x_{+}=mr_{0}e^{\theta}e^{-\hat{\chi}}$ is a dimension preserving
parameter which (after having taken the limit) has lost its physical
significance of a mass gap (the physical mass is the gap in the $P^{2}%
=P_{-}P_{+}$ spectrum).

Since this limit only effects the numerical factors and not the Fock space
operators $a^{^{\#}}(\theta),$ we expect that there will be no problem with
the horizontal restriction i.e. that the formal method (the last line in
\ref{lim}) agrees with the more rigorous result using smearing functions. Up
to a fine point which is related to the well-known infrared behavior of a
scalar chiral $dimA=0$ field, this is indeed the case. Using the limiting
$\chi$-parametrization we see that for the smeared field with $supp\tilde
{f}\in W,\,\tilde{f}$ real, one has the identity
\begin{align}
&  \int A(x_{+},x_{-})\tilde{f}(x)d^{2}x=\int_{C}a(\theta)f(\theta)=\int
A_{LF}(x_{+})\tilde{g}(x_{+})dx_{+},\,\tilde{g}\text{ }real\label{rap}\\
&  \tilde{f}(x)=\int_{C}e^{ip(\theta)x}f(\theta)d\theta,\,\,\;\tilde{g}%
(x_{+})=\int e^{ipx_{+}}g(p)\frac{dp}{\left|  p\right|  }=\int_{C}%
f(\theta)e^{ip_{-}(\theta)x_{+}}d\theta\nonumber
\end{align}
These formulas, in which a contour $C$ appears, require some explanation. The
onshell character of free fields restricts the Fourier transformed test
function to their mass shell values with the backward mass shell corresponding
to the rapidity on the real line shifted downward by $-i\pi$
\[
f(p)|_{p^{2}=m^{2}}=\left\{
\begin{array}
[c]{c}%
f(\theta),\,\,p_{0}>0\\
f(\theta-i\pi),.\,\,\,p_{0}<0
\end{array}
\right.
\]
and the wedge support property is equivalent to the analyticity of $f(z)$ in
the strip -$i\pi<\operatorname{Im}z<0$. The integration path $C$ consists of
the upper and lower rim of this strip and corresponds to the negative/positive
frequency part of the Fourier transform. By introducing the test function
$\tilde{g}(x_{+})$ which is supported on the halfline $x_{+}\geq0,$ it becomes
manifest that the smeared field on the horizon rewritten in terms of the
original Fourier transforms must vanish at $p=0$ as required by $L_{1}%
$-integrability according to
\begin{align}
f(p)|_{p^{2}=m^{2},p_{0}>0} &  \frac{dp}{\sqrt{p^{2}+m^{2}}}=f(\theta
)d\theta\equiv g(\theta)d\theta=g(p)|_{p^{2}=0,p_{0}>0}\frac{dp}{\left|
p\right|  }\\
&  \curvearrowright g(p=0)=0\nonumber
\end{align}
with a similar formula for negative $p_{0}$ and the corresponding $\theta
$-values at the lower rim. This infrared restriction is typical for spinless
free fields with $dimA=0.$ The equality of the $f$-smeared $A(x)$ field with a
$g$-smeared $A_{+}(x_{+})$ leads to the vanishing of $g(p)$ at the origin and
finally to the equality of the Weyl operators and hence of the generated
operator algebras
\begin{align}
\mathcal{A}(W) &  =\mathcal{A}(R_{+})\label{3}\\
\mathcal{A}(W) &  =alg\left\{  e^{iA(f)}|\,supp\tilde{f}\subset W\right\}
\nonumber\\
\mathcal{A}(R_{+}) &  =alg\left\{  e^{iA_{LF}(g)}|\,supp\tilde{g}\subset
R_{+},\int\tilde{g}dx_{+}=0\right\}  \nonumber
\end{align}
The choice of the lower causal horizon of $W$ would have led to the same
global algebra and could also serve as the starting point of LFH. The equality
(\ref{3}) is the quantum version of the classical causal shadow propagation
property of characteristic data on the upper lightfront of a wedge. With the
exception of $d=1+1,\,m=0$ where one needs the data on both light rays$,$ the
classical amplitudes inside the causal shadow $W$ of $R_{+}$ are uniquely
determined by either the upper or the lower data.

As in the classical analog of propagation of characteristic data, it would be
incorrect to think that the global identity of the algebras persists on the
local level i.e. that there exists a region in $W$ whose associated operator
algebra corresponds to a $\mathcal{A}(I)$ algebra when $I\in R_{+}$ is a
finite interval. The spacetime net structures on $W$ and $R_{+}$ are very
different; the subalgebras localized in finite intervals $\mathcal{A}(I)$ have
no local relation to localized subalgebras $\mathcal{A}(\mathcal{O})$ of
$\mathcal{A}(W)$ and vice versa; rather the position of compactly localized
algebras in one net is diffuse (``fuzzy'') relative to the net structure of
the other net. The fact that a finite interval on $R_{+}$ does not cast a
2-dimensional causal shadow does of course not come as surprise, since even in
the classical setting a causal shadow is only generated by characteristic data
which have at least a semi-infinite extension to lightlike infinity. Related
to this is the fact that the opposite lightray translation
\begin{align}
&  AdU_{-}(a)\mathcal{A}(R_{+})\subset A(R_{+})\\
&  U_{-}(a)=e^{-iP_{-}a}\nonumber
\end{align}
is a totally ``fuzzy'' endomorphism of the $A(R_{+})\,\,$net; whereas in the
setting of the spacetime indexing in the$\mathcal{A}(W)$ net, the map
$AdU_{-}(a)$ permits a geometric interpretation.

It is very important to notice that even in the free case the horizontal limit
is conceptually different from the scale invariant massless limit. The latter
cannot be performed in the same Hilbert space since the $m\rightarrow0$ limit
needs a compensating $\ln m$ term in the momentum space rapidity $\theta$ in
the argument of the operators $a^{\#}(\theta),$ whereas the horizon limit
(\ref{lim}) was only taking place in the c-number factors.

There is however no problem of taking this massless limit in correlation
functions if one uses the appropriate spacetime smearing functions. The
limiting correlation functions define via the GNS construction of operator
algebras a new Hilbert space \cite{Haag} which contains two chiral copies of
the conformal $dimA=0$ field corresponding to the right/left movers. This
difference between the scaling limit and the lightray holography limit for
free fields is easily overlooked, since the conformal dimensions of the
resulting fields are in this case the same and the only difference is that the
scaling limit leads to a 2-dimensional conformal theory which decomposes into
two independent chiral algebras. We will see that in the case of interacting
2-dimensional theories the appropriately defined holographic projection onto
the lightray is different from the scaling limit by much more than just a doubling.

There arises the question whether the chiral theories originating from the
lightray restriction are intrinsically different from those which are obtained
by factorizing d=1+1 conformal theories into its two chiral components. At
least in the present example this is not the case; we can always extend a free
chiral field independent of its origin to a d=1+1 massive field by defining an
additional action $U_{-}(a)$ which just creates a phase factor $e^{ip_{+}%
x_{-}}$ on the $a^{\#}(p)$ appearing in (\ref{lim}) without enlarging the
Hilbert space. The situation is reminiscent of Wigner's finite helicity
massless representations which allow an (in that case locally acting)
extension from the Poincar\'{e}- to the conformal- symmetry without
enlargement of the one-particle space.

Passing now to the higher dimensional case we notice, that by introducing an
effective mass which incorporates the transverse degrees of freedom on the
upper lightfront horizon of $W,$ the previous arguments continue to hold
\begin{align}
A(x) &  =\frac{1}{\left(  2\pi\right)  ^{\frac{3}{2}}}\int\left(
e^{-ip_{+}x_{-}-ip_{-}x_{+}+ip_{\perp}x_{\perp}}a(p)+h.c.\right)  \frac
{d^{3}p}{2\omega},\,\,x_{\pm}=x^{0}\pm x^{1}\\
p &  =(m_{eff}ch\theta,m_{eff}sh\theta,p_{\perp}),\text{\thinspace\thinspace
}m_{eff}=\sqrt{m^{2}+p_{\perp}^{2}},\,\,p_{\pm}=\frac{p^{0}\pm p^{1}}%
{2}\nonumber
\end{align}
The limiting field can again be written in terms of the same Fock space
creation/annihilation operators and as before the (effective) mass looses its
physical role
\begin{align}
A_{LF}(x_{+},x_{\perp}) &  =\frac{1}{\left(  2\pi\right)  ^{\frac{3}{2}}}%
\int\left(  e^{-ip_{-}x_{+}+ip_{\perp}x_{\perp}}a(p)+h.c.\right)  \frac
{dp_{-}}{2\left|  p_{-}\right|  }d^{2}p_{\perp}\\
A_{LF}(gf_{\perp}) &  =\int a^{\ast}(p_{-},p_{\perp})g(p_{-})f_{\perp
}(x_{\perp})\frac{dp_{-}}{2\left|  p_{-}\right|  }d^{2}p_{\perp}+h.c.\nonumber
\end{align}
In the second line the resulting operator is written in terms of a dense set
of test functions which factorize in a longitudinal (along the lightray) and a
transverse part. The longitudinal integration may be again brought into the
rapidity space form (\ref{rap}) involving a path $C.$ The dependence of the
longitudinal part on the transverse momenta is concentrated in the effective
mass, which in turn only enters via a numerical dimension-preserving factor in
$p_{-}x_{+}=m_{eff}e^{\theta}e^{-\chi}.$ Note that a description of the field
in the wedge in terms of lightray coordinates $x_{\pm}$ would have led to the
transverse dependence to be converted into the mass distribution of a
generalized field which is less simple than the lightfront projection.

In fact the best way to formulate the resulting structure on the lightfront is
to say that the longitudinal structure is that of a chiral QFT (with the
typical vacuum polarization leading to long range correlations) whereas
transversely it is quantum mechanical i.e. free of vacuum polarization and
without coupling between transverse separated subsystems. The ensuing
correlations are given by the inner product
\begin{align}
&  \left\langle A_{LF}(gf_{\perp})A_{LF}(g^{\prime}f_{\perp}^{\prime
})\right\rangle =\int\bar{g}(p)g^{\prime}(p)\frac{dp}{2\left|  p\right|  }%
\int\bar{f}_{\perp}(p_{\perp})f_{\perp}^{\prime}(p_{\perp})d^{2}p_{\perp
}\nonumber\\
&  \left[  A_{LF}(x_{+},x_{\perp}),A_{LF}(x_{+}^{\prime},x_{\perp}^{\prime
})\right]  =i\Delta(x_{+}-x_{+}^{\prime})_{m=0}\delta(x_{\perp}-x_{\perp
}^{\prime}) \label{op}%
\end{align}
where the second line shows that the commutation structure of the transverse
part is like that of \ Schr\"{o}dinger field. In fact the analogy to QM is
even stronger since the vacuum does not carry any transverse correlation at
all, a fact which can be best seen in the behavior of the Weyl generators
\begin{align}
\left\langle W(g,f_{\perp})W(g^{\prime},f_{\perp}^{\prime})\right\rangle  &
=\left\langle W(g,f_{\perp})\right\rangle \left\langle W(g^{\prime},f_{\perp
}^{\prime})\right\rangle \text{ }if\,\,suppf\cap suppf^{\prime}=\emptyset
\label{Weyl}\\
W(g,f_{\perp})  &  =e^{iA_{LF}(gf_{\perp})}\nonumber
\end{align}
i.e. the vacuum behaves like a quantum mechanical vacuum with no correlations
in the transverse direction\footnote{The lightfront restriction amounts to a
global change of the free field operators so that even spacelike correlations
on the lightfront become modified compared with their old value in the ambient
theory.}. To make this relation with transverse QM complete, we will now show
that the loss of vacuum correlations is accompanied by the appearance of a
Galilei group acting on these transverse degrees of freedom.

For this purpose it is helpful to understand the symmetry group of the
lightfront restriction. It is not difficult to see that it consists of a
7-parametric subgroup of the 10-parametric Poincar\'{e} group; besides the
longitudinal lightray translation and the $W$-preserving L-boost (which
becomes a dilation in lightray direction on the lightfront) there are two
transverse translation and one transverse rotation. The remaining two
transformations are harder to see; they are the two ``translations'' in the
Wigner little group (invariance group of the lightray in the lightfront). We
remind the reader that the full little group is isomorphic to the double
covering of the 3-parametric Euclidean $E(2)$ group in two dimensions. As a
subgroup of the 6-parametric Lorentz group it consists of a rotation around
the spatial projection of the lightray and two little group ``translations''
which turn out to be specially tuned combinations of L-boosts and rotations
which tilt the edge of the wedge in such a way that it stays inside the
lightfront but changes its angle with the lightray. This 2-parametric abelian
subgroup of ``translations'' corresponds in the covering $SL(2,C)$ description
of the Lorentz group to the matrix
\begin{align}
&  \left(
\begin{array}
[c]{cc}%
1 & a_{1}+ia_{2}\\
0 & 1
\end{array}
\right)  \\
G_{i} &  =\frac{1}{\sqrt{2}}(M_{it}+M_{iz}),\,\,i=x,y\nonumber
\end{align}
where in the second line we have written the generators in terms of the
Lorentz-generators $M_{\mu\nu}$ so that the above interpretation in terms of a
combination of boosts and rotations is obvious. The velocity parameter of the
Galilei transformations in the $x_{\perp}$-$x_{+}$ variables in terms of the
$a_{i},i=1,2$ can be obtained from the $SL(2,C)$ formalism.

The important role of this Galilei group in the partial return to quantum
mechanics as \textit{the} main simplifying aspect of \ LFH cannot be
overestimated. In the algebraic LFH see below) the lightray translation and
dilation set the longitudinal net structure whereas the Galilei
transformations are indispensable for re-creating the transverse localization
structure which was lost by the global identification of the wedge- with the
LFH-algebra (\ref{3}).

Note that (as in the 2-dimensional example) the physical particle spectrum is
not yet determined by these 7 generators; one rather needs to know the action
of the $x_{-}$ lightlike generator normal to the lightfront in order to obtain
the physical mass operator of the ambient theory (which is part of inverse LFH).

For free massless Bose fields (as for certain more general chiral fields)
there is no problem to cast the above pointlike formalism into the setting of
bounded operator algebras by either using the spectral theory of selfadjoint
unbounded operators or by Weyl-like exponentiation (see below). For free Fermi
fields the test function smearing suffices to convert them into bounded
operators; in this case one can elevate the above observations on smeared
fields directly into properties of spacetime-indexed nets of operator algebras.

There exists however a disappointing limitation for this lightfront
restriction formalism for pointlike interacting fields which forces one to
adopt the operator-algebraic method. Namely this restriction method suffers
from the same shortcomings which already affected the canonical equal time
formalism: with the exception of some superrenormalizable interactions in low
dimensional spacetime, there are no interacting theories which permit a
restriction to the lightfront or to equal times. In particular fields of
strictly renormalizable type (to which all Lagrangian fields used in d=1+3
particle physics belong) are outside the range of the above restriction
formalism. In fact a necessary condition for a restriction can be abstracted
from the two-point function and consists in the finiteness of the wave
function renormalization constants $Z$ which in the non-perturbative setting
amounts to the convergence of the following integral over the well-known
Kallen-Lehmann spectral function
\begin{align}
&  Z\simeq\int\rho(\kappa^{2})d\kappa^{2}<\infty\label{finite}\\
&  \left\langle A(x)A(0)\right\rangle =\int i\Delta^{(+)}(x,\kappa^{2}%
)\rho(\kappa^{2})d\kappa^{2}\nonumber
\end{align}
The operator algebra approach overcomes this restriction by bypassing the
short-distance problem of pointlike field ``coordinatization''. It uses the
\textit{causal shadow property} i.e. the requirement that the (weakly closed)
operator algebra associated with a simply connected convex spacetime region is
equal to the algebra of its causal completion\footnote{The extension by
spacelike ``caps'' of a timelike interval is already a consequence of the
spectrum condition and spacelike commutativity.} (causal shadow). In this way
certain operator subalgebras in the net of $W$ become identified with some
subalgebras localized on the horizon. For example a 3-dim. lightlike
semiinfinite strip within the horizon $H_{W}$ casts a causal shadow into $W$
which is simply the 4-dim. causal completion (the 4-dim. slab which this
3-dim. strip cuts into the ambient space). The largest such region is of
course the causal horizon $H_{W}$ itself, whose causal shadow is $W$
(\ref{3})$.$ The net on the lightfront also contains regions with a finite
longitudinal extension which do not cast causal shadows (independent of
whether their transverse extension is compact or reaches to infinity) because
they are already identical to their causal shadows. The algebras of those
regions do not correspond to algebras in the ambient net; they have to be
constructed by modular methods (using modular inclusions and modular
intersections) which will be presented in the next section.

The fact that the canonical equal time and lightfront restriction approach
suffer the same limitations for their Kallen-Lehmann spectral function does
not mean that their algebraic structure is similar. In fact it is well-known
that there are no pointlike generators for interacting canonical equal time
algebras for fields with infinite wave function renormalization, whereas (as
will be seen in the sequel) lightfront algebras permit a description in terms
of (generalized) chiral fields which allow for arbitrary noncanonical short
distance behavior. But the failure of the restriction procedure prevents their
\textit{direct} construction, rather one needs the algebraic detour of LFH to
get to those fields (in their role as locally coordinatizing the algebras).

Despite the totally different nature of the modular LFH method in the
formulation of an algebraic lightfront holography of the next section, most of
the structural results turn out to be analogous to those of the restriction
method in this section. Considered as a QFT in its own right, the LFH has some
unusual properties which are a consequence of the fact that the lightfront
does not belong to the family of globally hyperbolic spacetime manifolds to
which one usually restricts field theoretic considerations; but precisely this
physical shortcoming makes them useful auxiliary tools in QFT. Not only do
longitudinal compact regions not cast any causal shadows into the ambient
Minkowski spacetime, there are even no such shadows (and a related Cauchy
propagation) which extend the region \textit{inside} the lightfront. The
related transverse quantum mechanical behavior without vacuum polarization has
the consequence that algebras, whose transverse localizations do not overlap
admit a tensor factorization (just like the inside/outside tensor
factorization in Schroedinger theory in the multiplicative second quantization
formulation\footnote{In the standard quantum mechanical formulation the
statistical independence between inside/outside would corresponds to the
additive decomposition of the corresponding Hilbert spaces.}with no vacuum
entanglement). As a consequence, the application of such subalgebras with
finite transverse extension to the vacuum does not lead to dense subspaces of
the Fock space, rather their closures defines genuinely smaller subspaces
(breakdown of the Reeh-Schlieder property). In fact this LFH is the only known
mechanism by which one encounters a partial return to quantum mechanics within
the setting of QFT without invoking any nonrelativistic approximation, as will
be shown in the sequel.

The crucial operator-algebraic property which is responsible for this somewhat
unexpected state of affairs is the existence of positive lightlike
translations\footnote{This argument is analogous to the proof that certain
algebras have translational invariant centers \cite{Borchers} as in the case
of the proof of the factorial property of wedge algebras \cite{JMP}. The loss
of correlations results from the fact that lightlike translation act in a
two-sided way on strips \cite{DriesslerI}.} $U_{e_{+}}(a)$ in (two-sided)
lightlike strip (whose causal shadows are lightlike slabs) algebra
$\mathcal{A}(l$-$strip)$. Let $A\in\mathcal{A}(l$-$strip)$ and $A^{\prime}%
\in\mathcal{A}(l$-$strip)^{\prime}.$ Since $U_{e_{+}}(a)$ acts on both the
algebra and its commutant (associated with the complement of the strip within
the lightfront) as an automorphism and leaves the vacuum invariant, we obtain
for the vacuum expectation values
\begin{equation}
\left\langle 0\left|  AU_{e_{+}}(a)A^{\prime}\right|  0\right\rangle
=\left\langle 0\left|  A^{\prime}U_{e_{+}}(a)^{\ast}A\right|  0\right\rangle
\end{equation}
But according to the positivity of the translations this requires the function
to be a boundary value of an analytic function which is holomorphic in the
upper as well as the lower halfplane. Due to the boundedness the application
of Liouville's theorem yields the constancy in $a$. The cluster property
(which in the weak form as it is needed here also applies to infinite
lightlike separations) then leads to
\begin{equation}
\left\langle 0\left|  AA^{\prime}\right|  0\right\rangle =\left\langle
0\left|  A\right|  0\right\rangle \left\langle 0\left|  A^{\prime}\right|
0\right\rangle
\end{equation}
which is the desired tensor factorization without entanglement of the vacuum.
By successive application of this argument to a strip-subalgebra of the
commutant, the lightfront algebra can be made to factorize into an arbitrary
number of nonoverlapping strip algebras. In fact for those lightfront algebras
which are associated with the restriction of free field, the two-sided strip
algebras can easily be seen to be type $I_{\infty}$ factors i.e. the full
lightfront algebra (which is equal to the global ambient algebra)
tensor-factorizes into full strip algebras
\begin{align}
\mathcal{A}(LF) &  =\mathcal{A}(M^{(3,1)})=B(\mathcal{H})\\
&  =\bigotimes_{i}\mathcal{A}(LF_{i})=\bigotimes_{i}B(H_{i})\nonumber\\
\mathcal{H} &  =\bigotimes_{i}\mathcal{H}_{i}\nonumber
\end{align}
Algebras with smaller longitudinal localizations are imbedded as unique
hyperfinite type III$_{1}$ factors in suitable two-sided strip algebras
$B(H_{i}).\,$Subalgebras with a semi-infinite or finite longitudinal extension
which are associated with non-overlapping strip algebras inherit this
factorization; in fact they fulfill a Reeh-Schlieder theorem within an
appropriate factor space. The transverse correlation-free factorization for
arbitrarily small strips in lightray direction suggests that the lightfront
algebras are generated in terms of pointlike fields of the form (\ref{op})
where in the interacting case the zero mass free fields should be replaced by
generalized chiral fields with (half)integer scale dimensions (see next
section). The arguments leading to the transverse factorization are not
affected by interactions and therefore it is helpful to collect the result in
form of a structural theorem of operator (von Neumann) algebras

\begin{theorem}
A subalgebra $\mathcal{A}$ of $B(H)$ admitting a two-sided lightlike
translation with positive generator is of type I, i.e. it tensor-factorizes
$B(H)$ as $B(H)=\mathcal{A}\otimes\mathcal{A}^{\prime}$
\end{theorem}

The two-sidedness of the lightlike positive energy translations is important
in the above argument; if they act only one-sided (i.e. as an algebraic
endomorphism) as e.g. in the case of the wedge algebras, one can only conclude
that the center is translation-invariant and agrees with the center of the
full algebra; in this case there is no tensor-factorization and the algebras
turn out to be of the same kind as typical sharp localized algebras, namely
hyperfinite type III$_{1}$ factors\footnote{A detailed understanding of such
operator algebras beyond the fact that they have very different properties (no
minimal projectors corresponding to optimal measurements, no pure states) from
quantum mechanical algebras is not required in this paper.}\cite{JMP}.

With these remarks which (as shown in the next section) continue to apply in
the presence of interactions, we prepared the ground in favor of area laws for
transverse additive quantities (as e.g. a would be localization-entropy)
associated with a horizon. Later this problem will be considered in more detail.

The restriction of free fields to a causal horizon can also be carried out for
the more interesting case of the lower rotational symmetric causal horizon of
a (without loss of generality) symmetric double cone. Again the analogous
restriction limit (with the lower apex placed at the origin, $r_{+}=t+r$ plays
the role of $x_{+}$) maintains the creation and annihilation operator
structure and the Hilbert space and also leads to a conformal invariant limit
from which the original massive field may be reconstructed via the application
of suitable symmetries which lead away from the horizon into the ambient
spacetime. In this case there is however no geometric modular theory (no
Killing vector) for the massive free theory; nevertheless the massless
restriction to the horizon acquires a geometric modular group in form of a
subgroup of a double cone-preserving conformal transformation \cite{H-L}.
Similar to the LF situation, the algebra on the lower horizon $H_{\mathcal{C}%
}$ is equal to the double cone algebra (but with different subnet structure)
\begin{equation}
\mathcal{A}(H_{\mathcal{C}})=\mathcal{A}(\mathcal{C})
\end{equation}
This phenomenon which leads to an enlarged symmetry in the same Hilbert space
has been termed ``symmetry-enhancement'' \cite{Verch}. As a result of the
equality two algebras and the shared Hilbert space, one could also say that a
diffuse acting modular symmetry becomes geometric (a diffeomorphism) on the
horizon. Unfortunately it is presently not known how to derive this result for
double cones in the presence of interactions when the method of restricting
pointlike fields to the horizon breaks down \footnote{There is presently no
formulation of modular inclusions (see next section) for a diffuse acting
modular group which restricts to the smaller algebra as a diffuse compression
of the smaller region.}.

\section{Algebraic holography and modular localization}

The algebraic construction for interacting theories with trans-canonical
Kallen-Lehmann spectral functions starts from the position of a wedge algebra
$\mathcal{A}(W)$ within the algebra of all operators in Fock space
$B(\mathcal{H})$ and the action of the Poincar\'{e} group on $\mathcal{A}(W).$
It is based on the modular theory of operator algebras and adds two new
concepts: \textit{modular inclusions} and \textit{modular intersections}. The
intuitive idea of LFH consists in realizing that the causal shadow property
identifies certain semiinfinite algebras on the horizon with their ambient
4-dimensional causal shadows; starting from those one may construct a full net
of compactly localized subalgebras on the horizon in terms of relative
commutants and intersections among those algebras. Modular theory in the form
of inclusions and intersections is a mathematical tool which makes this
intuitive picture precise. Since these modular concepts have already received
attention in the recent literature on algebraic QFT, we will limit ourselves
to remind the reader of the relevant definitions and theorems (with a
formulation which suits our purpose) before commenting on them and applying
them to the lightfront holography.

\begin{definition}
(Wiesbrock, Borchers \cite{JMP}) An inclusion of operator algebras
($\mathcal{A}\subset\mathcal{B},\Omega$) is ''modular'' if ($\mathcal{A}%
$,$\Omega$), ($\mathcal{B}$,$\Omega$) are standard and $\Delta_{\mathcal{B}%
}^{it}$ acts (for t%
$<$%
0 by convention) as a compression on $\mathcal{A}$
\begin{equation}
Ad\Delta_{\mathcal{B}}^{it}\mathcal{A\subset A}%
\end{equation}
A modular inclusion is standard if the relative commutant ($\mathcal{A}%
^{\prime}\cap\mathcal{B},\Omega$) is standard. \ If the sign of t for the
compression is opposite it is advisable to add this sign and talk about a
$\pm$modular inclusion.
\end{definition}

Modular inclusions are different from the better known inclusions which arise
in the DHR superselection theory \cite{Haag} associated with the origin of
internal symmetries in quantum field theory. The latter are characterized by
the fact that they possess conditional expectations \cite{Lo-Re}. The
prototype of a conditional expectation in the conventional formulation of QFT
in terms of charge carrying fields is the projection in terms of averaging
over the compact internal symmetry group with its normalized Haar measure. If
$U(g)$ denotes the representation of the internal symmetry group we have
\begin{align}
\mathcal{A}  &  =\int AdU(g)\mathcal{F}d\mu(g)\\
E  &  :\mathcal{F}\overset{\mu}{\longrightarrow}\mathcal{A}\nonumber
\end{align}
i.e. the conditional expectation $E$ projects the (charged) field algebra
$\mathcal{F}$ onto the (neutral) observable algebra $\mathcal{A}.$

Modular inclusions have no conditional expectations. This is the consequence
of a theorem of Takesaki \cite{Tak} which states that the existence of a
conditional expectation for an inclusion between two
noncommutative\footnote{In the commutative case there are no restrictions.
Examples of commutative conditional expectations are the Kadanoff-Wilson
renormalization-group decimation procedures in the Euclidean setting of QFT.}
algebras (in standard position with respect to the same vector) is equivalent
to the modular group of the smaller being the restriction of that of the
bigger algebra. Since a genuine one-sided modular compression (endomorphism)
excludes the modular group of the smaller being the restriction of that of the
bigger algebra, there can be no projection $E$ in this case$.$ The modular
inclusion situation may be considered as a \textit{generalization} of the
situation covered by the Takesaki theorem.

The main aim of modular inclusion is to generate spacetime symmetry and nets
of spacetime indexed algebras which are covariant under these symmetries. From
the two modular groups $\Delta_{\mathcal{B}}^{it},\Delta_{\mathcal{A}}^{it}$
of a modular inclusion one can form the translation-dilation group with the
commutation relation $\Delta_{\mathcal{B}}^{it}U(a)=U(e^{-2\pi t}%
a)\Delta_{\mathcal{B}}^{it}$ and a system of local algebras obtained by
applying these symmetries to the relative commutant $\mathcal{A}^{\prime}%
\cap\mathcal{B}$ which may be combined into a possibly new algebra
$\mathcal{C}$%
\begin{equation}
\mathcal{C}\equiv\overline{\bigcup_{t}Ad\Delta_{\mathcal{B}}^{it}%
(\mathcal{A}^{\prime}\cap\mathcal{B})}%
\end{equation}
In general $H_{\mathcal{C}}\equiv\overline{\mathcal{C}\Omega}\mathcal{C}%
\subset H_{\mathcal{B}}=\mathcal{B\Omega\equiv}H,$ and whereas the modular
groups $\Delta_{\mathcal{B}}^{it},\Delta_{\mathcal{A}}^{it}$ of the inclusion
$\mathcal{A}\subset\mathcal{B}$ are different, the $\mathcal{C}\subset
\mathcal{B}$ inclusion leads to a Takesaki situation with $\Delta
_{\mathcal{C}}^{it}=\Delta_{\mathcal{B}}^{it}|_{H_{\mathcal{C}}}$ with the
conditional expectation being $E:\mathcal{B}\rightarrow\mathcal{C=}%
P\mathcal{B}P,\,H_{\mathcal{C}}=H_{\mathcal{B}}.$ If the inclusion is however
standard (which means $H_{\mathcal{C}}=H),$ the equality $\mathcal{C=B}$
follows. In that case a modular inclusion gives rise to a chiral structure on
$\mathcal{B}$ thanks to the following theorem.

\begin{theorem}
(Guido,Longo and Wiesbrock \cite{GLW}) Standard modular inclusions are in
correspondence with strongly additive chiral AQFT
\end{theorem}

Here chiral AQFT is any net of local algebras indexed by the intervals on a
line with a Moebius-invariant vacuum vector and the terminology
\textit{strongly additive} refers to the fact that the removal of a point from
an interval does not change the algebra i.e. the von Neumann algebra generated
by the two pieces is still the original algebra. This raises the question of
whether the present use of the word chiral is the same as that in the standard
literature where chiral refers to the apparently more restrictive situation of
the existence of an energy-momentum tensor which generates the diffeomorphisms
of a circle with the Moebius group being the maximal symmetry group of the
vacuum. However there are arguments that not only the Moebius transformations
(whose modular origin has been known for some time \cite{JMP}) but also the
diffeomorphisms beyond are of a general modular origin (for multi-local
algebras with respect to states which differ from the vacuum \cite{L-S}%
\cite{Kurusch}) . Hence the Witt-Virasoro algebra formed by the infinitesimal
generators of $Diff(S^{1})$ (which is the hallmark of standard chiral models
and follows from the existence of an energy-momentum tensor) appears to be
shared by the more general looking algebraic definition. This is helpful in
connection with LFH where generalized chiral theories arise from higher
dimensional QFT for which the chiral energy-momentum tensor is void of any
direct physical meaning. .

First we adapt the abstract theorem to our concrete case of a wedge algebra in
a massive interacting QFT in d=1+1 spacetime dimensions.
\begin{align}
\mathcal{B}  &  =\mathcal{A}(W)\\
\mathcal{A}  &  =U(e_{+})\mathcal{A}(W)U^{\ast}(e_{+}),\,e_{+}%
=(1,1)\nonumber\\
&  \equiv\mathcal{A}(W_{e_{+}})\nonumber
\end{align}
As Wiesbrock has shown \cite{W93}, the modular nature of the inclusion is a
consequence of the \textit{modular covariance} for wedge algebras (the
Bisognano-Wichmann property \cite{Haag}) This is the inclusion of the algebra
translated via a lightlike translation into itself so that geometrically the
relative commutator
\begin{equation}
\mathcal{A}(W_{e_{+}})^{\prime}\cap\mathcal{A}(W)\equiv\mathcal{A}(I(0,1))
\end{equation}
is by causality localized in the upper horizontal interval $(0,1)$. The
standardness of this inclusion then leads to a chiral conformal AQFT, i.e. a
net (more precisely a pre-cosheaf \cite{GLW})
\begin{align}
I  &  \rightarrow\mathcal{A}(I),\,\,I\subset S^{1}\\
\mathcal{A}(R_{+})  &  =\overline{\cup_{t}Ad\Delta^{it}\mathcal{A}%
(I(0,1))}\nonumber\\
\mathcal{A}(R)  &  =\mathcal{A}(R_{+})\vee J\mathcal{A}(R_{+})J\nonumber
\end{align}
on which the Moebius group (which preserves the vacuum vector) acts. With the
help of the external (i.e. non-Moebius) automorphism on $\mathcal{A}(R)$
implemented by the opposite lightray translation $U_{-}(a),$ we are able to
return from the chiral net on the right upper horizon to the original 2-dim.
net on $W$. If we call the transition from the d=1+1 original net via the
modular inclusion of wedge algebras to the $\mathcal{A}(R)$ net the
\textit{holographic projection}, then the reconstruction of the d=1+1 theory
from its holographic projection together with the opposite lightray
translation $U_{-}(a)$ (which acts as a kind of positive spectrum Hamiltonian)
should be called the \textit{holographic inversion}. Since chiral theories are
simpler than massive d=1+1 models, the gain by looking first at holographic
projections should be obvious. In fact the kind of chiral theory which is a
candidate for such a start is restricted to models with generating fields with
(half)integer scaling dimension which are closed under commutation i.e. where
the delta function terms and their derivatives are multiplied with field from
the generating set. Such models are ``Lie-fields'' in case of a finite number
of generators have been called ``W-algebras'' (here used in the wider sense
that current algebras are included).

In analogy to the free case of the previous section the $U_{-}(a)$ translation
does not enlarge the Hilbert space nor the global algebra on the horizon. But
without the knowledge of its action it would not be possible to identify the
physical mass via the mass operator $P_{+}P_{-}$ in that Hilbert space, nor
would one be able to recover the spacetime interpretation (e. i. the physical
content) of the original ambient theory. If the canonical quantization (or the
equivalent functional integral representation approach) would be an
ultraviolet consistent approach beyond canonical short distance behavior, then
the lightfront data acted upon by $P_{-}$ would be analogous to the canonical
data acted upon by the Hamiltonian. According to the remarks in the previous
section based on the Kallen-Lehmann representation, the canonical structure
gets lost in the process of renormalization, whereas the LFH-structure is
consistent with chiral fields of arbitrary high dimensions. This stability
against trans-canonical ultraviolet behavior favors the LFH approach as
compared to canonical method. Another potentially helpful aspect of the
constructive use of LFH is the fact that LFH data are capable to contain more
specific dynamical information since the set of chiral algebras is much richer
than the set of canonical data (which for given mass and spin is essentially
unique). It is generally believed that any change of the cut separating
dynamics from kinematics (which by definition contains the already
well-understood aspects) towards the enlargement of the latter is helpful in
the understanding and constructions of models of QFT.

An interesting family of models for which a study of these holographic aspects
seems to be well in reach, are the d=1+1 factorizing models for which some
modular localization properties are already known \cite{S}. Another advantage
of having a richer kinematical side may be that certain rather general
structural properties may become more accessible. In fact we will argue in the
next section that the area proportionality of localization entropy associated
with the wedge horizon in d=1+3 QFT is among those properties.

For the extension of holographic projection to higher dimensional theory one
needs one more mathematical definition and a related theorem about
\textit{modular intersections}.

\begin{definition}
(\cite{Wies}) A ($\pm$) modular intersection is defined in terms of two
standard pairs ($\mathcal{N}$,$\Omega$),$\,(\mathcal{M},\Omega)$ whose
intersection is also standard ($\mathcal{N\cap M},\Omega$) and which in
addition fulfill
\begin{align}
&  \left(  \left(  \mathcal{N\cap M}\right)  \subset\mathcal{N}%
,\mathcal{\Omega}\right)  \text{ }and\text{ }\left(  \left(  \mathcal{N\cap
M}\right)  \subset\mathcal{M},\mathcal{\Omega}\right)  \,\,is\,\ \pm
\,\operatorname{mod}ular\label{int}\\
&  J_{\mathcal{N}}(\lim_{t\rightarrow\mp\infty}\Delta_{\mathcal{N}}^{it}%
\Delta_{\mathcal{M}}^{-it})J_{\mathcal{N}}=\lim_{t\rightarrow\mp\infty}%
\Delta_{\mathcal{M}}^{it}\Delta_{\mathcal{N}}^{-it}=J_{\mathcal{M}}%
\lim_{t\rightarrow\mp\infty}\Delta_{\mathcal{N}}^{it}\Delta_{\mathcal{M}%
}^{-it})J_{\mathcal{M}}\nonumber
\end{align}
All limits are in the modular setting are to be understood in the sense of
strong convergence on Hilbert space vectors.
\end{definition}

We will now explain how this definition is used in LFH in order to achieve a
transverse localization on the lightfront. Setting $\Omega=$vacuum,
$\mathcal{M=A(}W\mathcal{)}$ and $\mathcal{N}=AdU(\Lambda_{e_{+}%
}(1))\mathcal{M}$ \ with $\Lambda_{e_{+}}(a)$ denotes a ``translation'' in the
Wigner little group which fixes the lightray vector $e_{+}$ (a transverse
Galilei transformation in the lightfront according to the previous section),
we first check the prerequisites of the definition. The modularity of the
inclusion follows again from the geometric properties of the action of the
modular group of the bigger algebra as a compression on the smaller and since
the limit in the second line is (\ref{int}) is nothing else but $U$%
($\Lambda_{e_{+}}(1))$ and the commutation relation with $J_{\mathcal{N}%
,\mathcal{M}}$ with $U$($\Lambda_{e_{+}}(1))$ in the last line are geometric
relation in the extended Lorentz group, the intersection property holds in our
case. In d=3+1 there are two independent ``translations'' which maintain the
given lightray and implement a Galilei transformation inside the lightfront.
As in the free field reduction formalism of the previous section, there is a
seven-parametric subgroup of the Poincar\'{e} group which is a symmetry of the
LFH. Since the Galilei transformations change the transverse directions inside
the lightfront, they tilt the transverse strips in such a way that the
intersection of the original with a transformed strip is a compact 3-dim.
region on the lightfront. These intersections define a net structure on the
lightfront containing arbitrarily small regions in the longitudinal as well as
transverse sense. Whereas it appears relatively straightforward to extend a
given bosonic/fermionic chiral theory by transverse quantum mechanical degrees
of freedom in a covariant way with respect to the seven-parametric symmetry
group, the remaining three symmetries, notably the Hamiltonian-like lightray
translation away from the lightfront (which are important to re-install the
particle physics interpretation of the ambient theory), pose the more
challenging task of holographic inversion. It is presently not even clear
whether all generalized chiral theories appear as images in LFH. From the
experience with the canonical Hamiltonian formulation one would expect that
this step is highly non-unique, or to phrase it the other way around that the
holographic projection is a many-to-one universality class relation.

The notion of a universality class projections is well-known in connection
with the scale and conformally covariant short distance limit which is the
basis of critical phenomena. The holographic projection classes constitute a
quite different universality class relation between (massive) theories in any
spacetime dimensions $d\geq1+1$ and chiral theories in which only
(half)integer dimensions appear. Even if the higher dimensional interacting
models do possess conventional pointlike field generators, there will be in
general no direct relation between these fields and those which generate the
chiral algebras. The holographic relation rather involves a radical spacetime
reprocessing which can only be formulated in terms of changing the spacetime
net structure; there seems to be no way to express such a radical change in
terms of a (necessarily) nonlocal formula which relates pointlike fields.

Since the theorem about the transverse tensor factorization of lightlike
strips and their causal shadows (lightlike slabs) of the previous section
follows from a general theorem about strip algebras with a two-sided action of
a positive lightlike translation which made no reference to free fields, it
remains valid in the interacting situation.

It is reasonable to ask whether these factorizing chiral strip algebras really
do permit the introduction of pointlike generating fields. The results in
\cite{F-J}\cite{J} on standard chiral theories suggest strongly that
group-representation methods may also be sufficient for their construction in
the present case. In case of a finite number of generating fields these fields
should be of the form of generalized W-algebras (Lie-fields) i.e. their exists
a finite collection of generating fields $A_{LF}^{(i)}(x_{+},x_{\perp
}),i=1,2,....$with the following (anti)commutation relation%

\begin{equation}
\left[  A_{LF}^{(i)}(x_{+},x_{\perp}),A_{LF}^{(j)}(x_{+}^{\prime},x_{\perp
}^{\prime})\right]  =\left\{  \sum_{k=0}^{n(i,j)}B_{LF,k}^{(i,j)}%
(x_{+},x_{\perp})\delta^{(k)}(x_{+}-x_{+}^{\prime})\right\}  \delta(x_{\perp
}-x_{\perp}^{\prime})
\end{equation}
where the sum extends over chiral $\delta$-functions and their derivatives
multiplied with fields $B$ which consist of linear combinations of generating
fields $A$ with the same scale dimension which together with the dimension
carried by the (derivatives of) $\delta$-functions match the balance of scale
dimension of the left hand side. For readers who are familiar with chiral
conformal QFT these formula are straightforward transverse extensions of
W-algebra commutation with covariance properties under the seven-dimensional
symmetry group of lightfront. The longitudinal compactification adds the rigid
rotation symmetry of the circle. As in the case of standard W-algebras this
structure is easily extended to include Fermion fields.

The quantum mechanical behavior in the transverse direction is described by
the common transverse $\delta$-function. Besides the covariance under the
seven-parametric subgroup of the Poincar\'{e} group, the expectation values
fulfill the strong factorization
\begin{equation}
\left\langle CD\right\rangle =\left\langle C\right\rangle \left\langle
D\right\rangle
\end{equation}
where $C,D$ are products of generating fields so that the transverse
coordinates in the $C$-cluster are disjoint from those of the $D$-cluster.
Another way to describe lightfront fields which makes the quantum mechanical
transverse structure more explicit would be to use the notion of wave
functions on transverse coordinates which are chiral field-valued.

It should be mentioned that modular intersections which we used to establish
the Galilei invariance of the transverse quantum mechanics associated with LFH
also played an important role in the construction of 3- and higher-
dimensional AQFT starting from a finite set of wedge algebras \cite{Kaehler}.

\section{Area density of horizon-associated entropy}

The use of the LFH for a coarse classification of higher dimensional theories
with the aim of their construction via holographic inversion is an ambitious
new program. A more modest goal would consists in trying to pinpoint general
properties of the LFH projection which pertain to the LFH universality
classes. In the sequel we will argue that the area proportionality of
\textit{localization entropy} associated with the horizon of a wedge is such a property.

It has been known for a long time that the restriction of the vacuum state to
the wedge algebra leads to a thermal state with a fixed Hawking temperature.
To place this formal mathematical observation on a physical footing, Unruh
\cite{Unruh} has interpreted this effect as a physical phenomenon seen by a
uniformly accelerated observer with a fixed acceleration (creating the causal
wedge horizon). The Hawking-Unruh temperature arises from the fact that an
Unruh observer is immersed in a KMS thermal state for which the
wedge-preserving Lorentz-boost defines the dynamics (the ``Hamiltonian'').
There are however obstacles against directly associating a notion of entropy
with this thermal situation, the most obvious being the fact that the Lorentz
boost is not a trace-class operator. In this section we will argue that the
LFH method combined with the so-called split property applied to the chiral
theories on the lightfront allows to extract a (relative or unnormalized) area
density of localization entropy. Our arguments depend on the assumption that
the (logarithmic) increase of vacuum polarization at the boundary of
localization in chiral theories in the limit of sharp boundaries (vanishing
splitting distance) is universal, so that different chiral models only lead to
different numerical factors which determine finite ratios. Only in this case
an assignment of an additive measure of the cardinality of degrees of freedom
associated with the vacuum state restricted to a sharp localizes spacetime
region as a wedge or its horizon is conceivable. The determination of a
\textit{normalized} area densities would then hinge on the validity of
fundamental thermodynamic laws involving localization entropy, an issue which
is outside the scope of this paper.

Whereas our conclusions are still preliminary, the ideas by which we try to
relate transverse area density of localization entropy on the lightfront with
vacuum polarization at the boundaries of localization are interesting in their
own right and may turn out to be useful in more profound future studies of
this problem.

It is interesting to recall that Heisenberg discovered vacuum fluctuations
through surface effects. Phrased in a more modern setting of partial charges
defined by smeared Noether currents
\begin{equation}
Q(f_{R,\varepsilon}f_{T})=\int j_{0}(x)f_{R,\varepsilon}(\vec{x})f_{T}%
(x_{0})d^{4}x
\end{equation}
where the spatial test function equals one inside a sphere of radius $R$ and
vanishes outside $R+\varepsilon$ and the time smearing is a positive test
function symmetric around $x_{0}=0$ and with total integral equal to one. This
partial charge has the same commutation relation with other operators as the
total charge as long as these operators are localized inside a
(origin-centered) double cone of size $R.$ By use of the Kallen-Lehmann
representation one can calculate the charge fluctuation
\begin{equation}
\left\langle \Omega\left|  Q(f_{R,\varepsilon}f_{T})^{2}\right|
\Omega\right\rangle =\left\|  Q(f_{R,\varepsilon}f_{T})\Omega\right\|
^{2}\sim R^{2}c(\varepsilon)
\end{equation}
they are (for small $\varepsilon$) proportional to the surface of the sphere
with a numerical coefficient which diverges as a inverse power in
$\varepsilon.$ This is not in contradiction with the expected zero value of
the total charge in the vacuum state because the convergence for
$R\rightarrow\infty$ is in the sense of weak convergence\footnote{One can also
have strong convergence at the expense of a time smearing which depends on the
spatial smearing \cite{R}.}.

In the algebraic setting the consequences of vacuum polarization have more
dramatic structural manifestations. Their omnipresence even in the absence of
interactions is behind the loss of such quantum mechanical properties as the
existence of minimal projectors and even that of pure states for the sharply
localized algebras which are the basic objects of the algebraic approach. As
already mentioned before in terms of von Neumann types they are isomorphic to
the unique hyperfinite type III$_{1}$ von Neumann factor instead of the unique
type $I_{\infty}$ factor isomorphic to the algebra of all operators $B(H)$ in
an infinite dimensional Hilbert space $H$. In the multiplicative second
quantization formulation of Schroedinger quantum mechanics the spatial
division into an algebra inside and outside a spatial box would lead to a
tensor factorization of the two algebras. Interactions would correlate
(entangle) the inside with the outside of the box but not affect the tensor
product factorization. The nonrelativistic vacuum would remain a
(nonentangled) tensor product of the in and outside vacuum even in the
presence of interactions. Hence the restriction of the global vacuum to the
inside of the box would never create an impure state with thermal aspects. The
situation changes radically in the presence of field theoretic vacuum
fluctuations. In that case not only does the vacuum become an entangled state
in a tensor product description, but the very tensor product structure, which
is the basis of the entanglement, gets lost. Namely the global algebra is not
representable as a tensor product of the local algebra $\mathcal{A(C)}$ (with
say a double cone $\mathcal{C}$ being the analog of a nonrelativistic box) and
its commutant $\mathcal{A(C)}^{\prime}$ (equal to the causally disjoint
algebra $\mathcal{A(C}^{\prime})$ according to Haag duality), although it is
generated by both commuting factors: $B(H)=\mathcal{A(C)\vee A(C}^{\prime}).$
This radical change is consistent with the global vacuum turning into a KMS
state at the Hawking temperature upon restriction from the global to a local
algebra; however the von Neumann entropy requires type I and cannot be defined
in such a situation. In order to obtain a tensor factorization one invokes the
``split property'' (for a rather complete bibliography see \cite{Haag}) which
is a kind of algebraic analog of the above test function transition region of
size $\varepsilon$ and intuitively speaking corresponds to a ``fuzzy'' surface
of a spacetime localization region. Algebraically one takes instead of sharply
localized algebra $\mathcal{A(C)}$ \ a fuzzy-localized type I algebra whose
localization is between $\mathcal{C}$ and the by $\varepsilon$ bigger double
cone $\mathcal{C}_{\varepsilon}.$ Let us first recall the general algebraic
definition without reference to our geometric setting.

\begin{definition}
\bigskip An inclusion $\mathcal{A}\subset\mathcal{B}$ is called split if there
exists an intermediate type I factor $\mathcal{N}$ : $\mathcal{A}%
\subset\mathcal{N}\subset\mathcal{B}$. This yields the following tensor
factorization: $B(H)=\mathcal{N}\otimes\mathcal{N}^{\prime}%
,\,\,\mathcal{A\rightarrow}\mathcal{A}\mathcal{\otimes1,\,B}^{\prime
}\rightarrow1\otimes\mathcal{B}^{\prime}$
\end{definition}

We remind the reader that in theories with reasonable phase space properties
there exists a canonical way of constructing for a standard inclusion (i.e.
$\left(  \mathcal{A}^{\prime}\cap\mathcal{B},\Omega\right)  $ is also
standard) a type I factor \cite{Haag} which is explicitly given by the
formula
\begin{align}
&  \mathcal{N}=\mathcal{A}J\mathcal{A}J=\mathcal{B}J\mathcal{B}J\\
&  with\,\,J\equiv J_{\left(  \mathcal{A}^{\prime}\cap\mathcal{B}%
,\Omega\right)  }\nonumber
\end{align}
\ The \ verification of these structural properties is not difficult
\cite{Do-Lo}, but the explicit computation of a modular conjugation $J$ which
belongs to not simply connected or disconnected localization region has not
been possible up to now \footnote{There has been some recent progress in the
understanding of modular objects associated with double interval chiral
algebras which may turn out to be helpful in future attempts. \cite{L-S}%
\cite{Kurusch}.}. The fact that $\mathcal{N}$ is type I implies that the
modular group of $\left(  \mathcal{N},\Omega\right)  $ is implemented by the
unitary group $h^{it}=e^{it\mathbf{H}_{\mathcal{N}}}$ with a generator
$\mathbf{H}_{\mathcal{N}}$ which is affiliated with the algebra $\mathcal{N}$
\begin{align}
&  Ad\Delta_{\left(  \mathcal{N},\Omega\right)  }^{it}\mathcal{N}%
=Adh^{it}\mathcal{N}\\
\Delta_{\left(  \mathcal{N},\Omega\right)  }^{it}H  &  =h^{it}H_{\mathcal{N}%
}\otimes h^{-it}H_{\mathcal{N}^{\prime}},\,\,H=H_{\mathcal{N}}\otimes
H_{\mathcal{N}^{\prime}}\nonumber
\end{align}
where the positive operator $h=h^{it}|_{t=-i}$ represents the
Connes-Radon-Nikodym derivative of the vacuum state $\omega_{\Omega}(\cdot)$
restricted to $\mathcal{N}$ with respect to the trace (tracial weight) on
$\mathcal{N}.$ By construction this operator has discrete spectrum and finite
trace $tr\,h<\infty$ (which by normalization $\omega_{\Omega}(\cdot
)=tr(h\cdot)$ can be set equal to one). The finiteness of the entropy
\begin{equation}
E_{\left(  \mathcal{N},\Omega\right)  }\equiv-tr|_{H_{\mathcal{N}}}hlnh
\end{equation}
requires in addition the absence of an infrared accumulation of too many
eigenvalues near zero.

Returning to the above geometric case, $\mathcal{A}$ and $\mathcal{B}$ should
be identified as $\mathcal{A=A(C)}$,$\,\mathcal{B=A(C}_{\varepsilon}$). The
main interest would be the study of the limiting $\varepsilon\rightarrow0$
behavior of the localization entropy $E_{\left(  \mathcal{N},\Omega\right)
}.\,$\ It is only in this limit that one would expect the localization entropy
to become independent of the choice of the splitting type I subfactor
$\mathcal{N}$ (similar to the dependence of the partial charge on the
$f_{R,\varepsilon}(\vec{x})$ test function dependence in the transition
region). In the algebraic setting one can actually avoid this problem of
arbitrariness in the splitting by realizing that all different choices of
$\mathcal{N}$ yield the same product state
\begin{equation}
\omega_{p}(AB^{\prime})=\omega(A)\omega(B^{\prime}),\,\,A\in\mathcal{A}%
,B^{\prime}\in\mathcal{B}^{\prime}%
\end{equation}
This together with Kosaki's elegant variational formula \cite{Kosaki}%
\cite{Narn} for Araki's relative entropy $E(\omega_{p},\omega)$ of the
original vacuum state $\omega$ with respect to the product vacuum would get
rid of the dependence of the special kind of tensor product implementation
even before going to the limit. In either formulation there have been attempts
to compute such localization entropies, but even in the case of algebras
associated with free fields the problem has turned out to be quite intractable
\cite{Detlev}\cite{Narn}

In the following we will argue that the issue of localization entropy for the
wedge can be significantly simplified by LFH. Although the wedge algebra is
identical to that of its horizon, the latter offers a better spacetime
arrangement of degrees of freedom for the computation of localization entropy.
The reason is obvious since the transverse symmetry together with strong
factorization property of the vacuum in transverse direction permit the
reduction of the entropy problem to that of an area density (entropy per unit
volume of the edge of the wedge) which is determined by the chiral theory on
the lightray. One then hopes to profit from the simplicity of chiral theories.

It is well-known that the split property of chiral theories is a consequence
of the nuclearity property which in turn follows from the convergence of the
partition function \cite{ALR}%

\begin{equation}
tre^{-\beta L_{0}}<\infty\label{trace}%
\end{equation}
where $L_{0}$ is the rigid rotation operator on the compactified light ray.
Since the transverse symmetry (which would lead to an infinite degeneracy of
these eigenvalues) has been taken out, this condition does not seem to be
unduly restrictive. After the transverse reduction the horizon now corresponds
to the right halfline or in the compactified circular description to the upper
semi-circle. The chiral split inclusion to be studied is now $\mathcal{A}%
\subset\mathcal{B}$ with $\mathcal{A}=\mathcal{A}(I(0+\varepsilon
,\pi-\varepsilon)),$ $\mathcal{B}=\mathcal{A}(I(0,\pi)).$ The practical
problem is then to find an implementation of the split isomorphism $\Phi$%

\begin{equation}
\mathcal{A}\vee\mathcal{B}^{\prime}\overset{\Phi}{\simeq}\mathcal{A}%
\otimes\mathcal{B}^{\prime}%
\end{equation}
which is computationally manageable. It turns out that implementation of
$\Phi$ with the help of the ``flip trick'' used in \cite{charges} is useful.
It consists in implementing $\Phi$ in the duplicated tensor representation
space $H\otimes H$ with the duplicated vacuum $\Omega\otimes\Omega.$ Assume
for simplicity that the algebra is generated by one pointlike field $\psi$ and
with $\psi_{1}\equiv\psi\otimes\mathbf{1,}$ $\psi_{2}\equiv\mathbf{1}%
\otimes\psi$ the implementation of $\Phi$ reads%

\begin{equation}
\Phi(\psi(x))=\left\{
\begin{array}
[c]{c}%
\psi_{1}(x),\,\,x\in I(0+\varepsilon,\pi-\varepsilon)\\
\psi_{2}(x),\,\;x\in I(\pi,2\pi)
\end{array}
\right. \nonumber
\end{equation}
If we now view $\psi_{i},i=1.2$ formally as a $SO(2)\simeq U(1)$ charge
doublet (each living in a separate tensor factor of $H\otimes H$), then we may
implement $\Phi$ in the spirit of the well-known Noether current formalism
\begin{align}
&  \Phi(\psi_{1}(x))=U(f)\psi_{1}(x)U(f)^{\ast},\,\,U(f)=e^{ij(f)}\label{ep}\\
j(f)  &  \equiv\int f(x)j(x)dx,\,\,f=\left\{
\begin{array}
[c]{c}%
0,\,\,\text{\thinspace}x\in I(0+\varepsilon,\pi-\varepsilon)\\
1,\,\,\,\,x\in I(\pi,2\pi)
\end{array}
\right. \nonumber
\end{align}
The bilinear expression for the current in terms of the two-component field
has a precise mathematical meaning in the case when the generating field is a
free chiral Boson or Fermion field\footnote{Iin the latter case the duplicated
Fermion formalism is related to the tensor product description by a Klein
transformation.}. This implementation of $\Phi$ by a unitary operator depends
again on the choice of the smearing function $f$ in the transition regions.
Before looking at the issue of localization entropy it is instructive to
consider the overlap between the tensor duplicated representation of the
original vacuum $\Omega_{vac}=\Omega\bar{\otimes}\Omega$ and the choice of
implementing vector $\Omega_{p}\equiv U(f)(\Omega\bar{\otimes}\Omega)$ of the
product vacuum%

\begin{align}
&  \left\langle \Omega_{vac}|\Omega_{p}\right\rangle =\left\langle
\Omega_{vac}\left|  U(f)\right|  \Omega_{vac}\right\rangle =e^{-\frac{1}%
{2}\left\langle j(f),j(f)\right\rangle _{0}}\\
&  \left\langle \eta^{\prime}\left|  AB^{\prime}\right|  \eta^{\prime
}\right\rangle =\left\langle \Omega_{vac}\left|  A\right|  \Omega
_{vac}\right\rangle \left\langle \Omega_{vac}\left|  B^{\prime}\right|
\Omega_{vac}\right\rangle \nonumber\\
&  A\in\mathcal{A}(I(0+\varepsilon,\pi-\varepsilon)),\,\,B^{\prime}%
\in\mathcal{A(}I(\pi,2\pi)\mathcal{)}\nonumber
\end{align}

The fluctuation of a smeared free current is easily shown to have a
logarithmic dependence on the $\varepsilon$ in (\ref{ep}) which leads to a
linear vanishing of the overlap in $\varepsilon.$ The vanishing of the overlap
is of course related to the expected unitary inequivalence of the product
representation in the limit $\varepsilon=0.$ In fact the vanishing of the
overlap can be shown to hold for all basis vectors in $H\otimes H$ which are
generated by polynomials in the creation operators%

\begin{equation}
\left\langle \Omega_{vac}\left|  U(f)\right|  a_{1}^{\ast}(p_{1}%
)...a_{n}^{\ast}(p_{n})\otimes a_{2}^{\ast}(k_{1})...a_{2}^{\ast}(k_{m}%
)\Omega_{vac}\right\rangle \rightarrow0
\end{equation}
By tracing the square of these matrix elements over the second tensor factor
basis, one obtains a density matrix $\rho_{\varepsilon}$ which represents the
vacuum $\Omega_{vac}$ as a mixed state on the factor space $H=\overline
{\mathcal{A(}I(0+\varepsilon,\pi-\varepsilon)\mathcal{)\Omega}_{p}}.$ The
associated von Neumann entropy%

\begin{equation}
s_{\varepsilon}=-\rho_{\varepsilon}ln\rho_{\varepsilon}=-ln\varepsilon\cdot s
\end{equation}
represents the desired localization entropy for this particular implementation
of the split localization. The implementation independent description which is
intrinsically determined in terms of product states is obtained by minimizing
over all implementations of the product state. The associated entropy is the
minimum of all split entropies and can be explicitly written in terms of the
aforementioned Kosaki variational formula.

The only presently computational accessible entropy is the one in the above
two fold replica tensor-factorized implementation for free chiral theories. We
hope to be able to present an explicit calculation along this scenario in a
future paper.

It is important to stress that the success of the LFH formalism to the problem
of area density of localization entropy depends on whether chiral theories
lead to a universal divergence for this density as the splitting distance
approaches zero. Only if this turns out to be true can one assign a split
independent additive area density to the horizon which is fixed by LFH up to a
common normalization constant (independent of the specific chiral model).
Under these circumstances the LFH analysis of Rindler wedges would lead to
well defined ratios between area densities of entropy. This (relative or
unnormalized) area density would be an intrinsic aspect of the LFH algebra
rather than (as in standard field theoretic attempts) a consequence of the
accidental short distance behavior of special field coordinatization.

It is interesting to note that this idea of a relative area density of
localization entropy would be in complete harmony with the already well
understood temperature aspect of localization on wedges or on their horizons
since the modular operator for the split situation converges against the
Lorentz-boost (which corresponds to the dilation in the LFH on the horizon)
representing the dynamics of the thermal KMS state. This suggests that the LFH
is capable to extract a physically meaningful measure of entropy area density
from the Lorentz boost by extracting an infinite area factor together with the
splitting dependence. Throughout the splitting process the restricted vacuum
remains a thermal KMS state (it is even a Gibbs state as long as $\varepsilon$
remains larger than zero) with the Hawking temperature An ad hoc cutoff (in
order to change the Lorentz boost into a trace class operator) in the short
distance regime of fields would not be physically reasonable since it is a
brute force method which wrecks the (local aspects of the) theory in an
uncontrollable manner which is totally unrelated to the Hawking temperature
aspects of localization. It is important to realize that despite its
appearance the splitting method (unlike cutoffs) does not destroy degrees of
freedom, it just redistributes them inside the boundary of thickness
$\varepsilon.$

\section{Conclusions and outlook}

By re-investigating the structure of lightfront QFT with the help of recent
concepts from local quantum physics we succeeded to give a concise meaning to
the idea of (algebraic) LFH. The most useful new result of the present LFH
formalism is the compression of vacuum fluctuations into the direction of the
lightray and a very strong form of quantum mechanical statistical independence
in transverse directions within the lightfront which manifests itself in the
absence of any transverse vacuum polarization. This appearance of quantum
mechanics within relativistic QFT (without having performed any
nonrelativistic approximation) is accompanied by the appearance of a genuine
transverse Galilei covariance which results from projecting the Wigner little
group ``translations'' into the lightfront.

The statistical independence of transverse separated subsystems has immediate
consequences for quantities which behave additively for uncoupled subsystems
as entropy. Namely the problem of assigning an entropy to the vacuum state
restricted to the horizon of a wedge is transferred to the question whether a
transverse area density of entropy can be introduced by restricting the vacuum
state of a global chiral theory to a local chiral subalgebra (for the case at
hand associated with half of the lightray).

It is well-known that the global vacuum becomes a thermal KMS state at the
Hawking temperature, but since the relevant generator of the KMS dynamics is
not given by a traceclass operator, the problem of finding a corresponding
entropy as a measure of the impurity of the vacuum caused by the restriction,
poses somewhat of a dilemma. Namely the same mechanism which causes the
impurity aspects (which are the prerequisites for a nontrivial entropy)
prevents the relevant dynamical operator to be of trace class. To overcome
this problem we inferred the split property of AQFT which replaces the KMS
state by an $\varepsilon$-dependent family of Gibbs states which for
$\varepsilon\rightarrow0$ approximates the KMS state. Different from a cutoff
implemented on the short distance fluctuation of distinguished pointlike
fields, this split procedure only re-shuffles the degrees of freedom within a
fuzzy interval of thickness $\varepsilon$ around the endpoints within the same
local chiral theory. Therefore contrary to the cutoff method it serves as an
excellent starting point for investigating a universal behavior in the limit
of $\varepsilon\rightarrow0.$ If further investigations of the split limit in
chiral theories confirm our conjecture of a universal (logarithmic) divergence
independent of the field content of the local algebras, then the LFH
universality classes (which were shown to be classified in terms of chiral
models) would only show up in numerical coefficients which are determined up
to a common model independent factor. In that case the distinction between
heat bath thermality and localization thermality, which presently manifests
itself in the fact that the heat bath temperature is variable whereas the
localization-caused temperature is determined by geometrical aspects, could be
enriched by a characteristic difference in their entropy behavior. Whereas the
heat bath situation leads to a volume density, the relevant concept for causal
localization is an area density. To link this to observable physics, the
process of localization which creates causal horizons should be taken out of
its abstract ``Gedanken'' setting in this paper and supplied with a concrete
physical blueprint in the spirit of Unruh \cite{Unruh}.

Note that the arguments in the present work hold for all causal quantum
theories with a maximal speed of causal propagation leading to a notation of
causal disjointness and local commutativity. This means in particular that all
(acoustical, optical, hydrodynamic) black hole analogs \cite{S-U} have a LFH.
The difference of gravitational interactions to their analogs lies in the
scale of couplings between the local quantum physics with geometric properties
(the analogs of ``surface gravity'').

The reader may ask why (despite obvious similarities) we refrained up to now
from relating our LFH based quantum results to the classical Bekenstein area
law. Of course we do not believe that the area law behavior is the result of
an accidental coincidence, but we think that this problem of comparison
requires more detailed and profound future studies, in particular a proof of
the conjectured universal behavior for vanishing split $\varepsilon
\rightarrow0.$ Local quantum theories have a much richer structure than their
classical counterparts. Whereas we find it entirely conceivable that the study
of fundamental dynamical laws for localization entropy in the context of black
holes and their analogs may lead to a normalized entropies (with the
normalization depending on the kind of analog), it is less plausible to expect
that all the quantum differences between LFH universality classes will be
lost\footnote{Apart from the classical ``surface gravity'' constant through
which the gravitational coupling enters into the Bekenstein formula.
Derivations of thermodynamic fundamental laws from classical field models can
be found in \cite{Wald}.} in favor of a universal classical Bekenstein law.

The results presented in this paper do not support the optimistic view that
the area density of entropy associated with the thermal aspect of restriction
to a horizon reveals more about the nature of quantum gravity than the Hawking
temperature and radiation. But while de-emphasizing the relation to a yet
unknown quantum theory of gravity they strengthen the (still somewhat
mysterious \cite{W-S}) links between thermal behavior and geometry.

\textit{Acknowledgments}: I am indebted to Jacob Yngvason for a invitation for
a visit of the ESI and I thank the organizers of the ESI workshop ``Quantum
field theory on curved space time'' for the opportunity to present the main
results of this paper.

\end{document}